%% file: SQDMsparseGP_CDC.tex
\documentclass[letterpaper, 10pt, conference]{ieeeconf}      % Use this line for a4 paper

\IEEEoverridecommandlockouts                              % This command is only needed if 
\usepackage{amsmath} 
\usepackage{textcase}
\usepackage{subfig}
\usepackage{amsfonts} 
\usepackage{amssymb}
\usepackage{bm}
\usepackage{mathrsfs} 
\usepackage{graphicx} 
\usepackage{xcolor}
\usepackage{tabularx}
\usepackage{units}
\usepackage{xspace}
\usepackage[noadjust]{cite}
\usepackage[font=footnotesize]{caption}
\usepackage{booktabs}			    % Changes table design (primarily the
\graphicspath{{figs/}}

\newcommand*{\ie}{i.e.,\xspace}
\newcommand*{\cf}{see\xspace}

\DeclareMathOperator*{\argmin}{arg\,min}

\newcommand*{\diff}{\mathop{}\!\mathrm{d}}	% Upright 'd' for differential
\newcommand*{\inv}{^{-1}}
\newcommand*{\nt}{\ensuremath{n_*}}			% Number of test inputs
\newcommand*{\nI}{\ensuremath{n_I}} 		% Number of active/inducing data
\newcommand*{\y}{\ensuremath{\gamma}}		% Symbol for GP training targets
\newcommand*{\R}{\ensuremath{\mathbb{R}}}
\newcommand*{\N}{\ensuremath{\mathcal{N}}}
\newcommand*{\GP}{\ensuremath{\mathcal{GP}}}
\newcommand*{\I}{\ensuremath{\mathbb{I}}}
\newcommand*{\CC}{\ensuremath{\mathcal{O}}}
\newcommand*{\Vneg}{\ensuremath{V^-}\xspace}
\newcommand*{\Vpos}{\ensuremath{V^+}\xspace}
\newcommand*{\Vmp}{\ensuremath{V^\mp}\xspace}

\newcommand*{\Vb}{\ensuremath{V_\text{b}}\xspace}

\newcommand*{\Df}{\ensuremath{\Delta f}\xspace}

\newcommand*{\SQDM}{scanning quantum dot microscopy\xspace}
\newcommand*{\EP}{electrostatic potential\xspace}
\newcommand*{\EPs}{electrostatic potentials\xspace}
\newcommand*{\DOF}{two-degree-of-freedom\xspace}
\newcommand*{\ESC}{extremum seeking controller\xspace}
\newcommand*{\Sn}{\ensuremath{\sigma^2_\text{n}}\xspace}

\renewcommand*{\mid}{\,|\,}

\title{\LARGE \bf
Fusing Online Gaussian Process-Based Learning and Control for Scanning
Quantum Dot Microscopy
}
\author{Maik Pfefferkorn$^{1}$, Michael Maiworm$^{1}$, Christian
Wagner$^{2,3}$, F. Stefan Tautz$^{2,3}$, Rolf Findeisen$^{1}$% <-this % stops a space
\thanks{$^{1}$Otto-von-Guericke-Universitaet Magdeburg, Laboratory for
        Systems Theory and Automatic Control, Germany, {\tt\small
        \{rolf.findeisen, michael.maiworm, maik.pfefferkorn\}@ovgu.de}.}%
\thanks{$^{2}$Peter Gruenberg Institute (PGI-3), Juelich Research
        Center, Juelich, Germany, {\tt\small c.wagner@fz-juelich.de}.}%
\thanks{$^{3}$Juelich Aachen Research Alliance (JARA) -- Fundamentals of
Future Information Technology, Juelich, Germany.}%
}

\begin{document}

\maketitle
\thispagestyle{empty}
\pagestyle{empty}

%%%%%%%%%%%%%%%%%%%%%%%%%%%%%%%%%%%%%%%%%%%%%%%%%%%%%%%%%%%%%%%%%%%%%%%%%%%%%%%%
\begin{abstract}
  Elucidating electrostatic surface potentials contributes to a deeper
  understanding of the nature of matter and its physicochemical
  properties, which is the basis for a wide field of applications. 
  Scanning quantum dot microscopy, a recently developed technique allows
  to measure such potentials with atomic resolution. 
  For an efficient deployment in scientific practice, however, it is
  essential to speed up the scanning process.
  To this end we employ a two-degree-of-freedom control paradigm, in which
  a Gaussian process is used as the feedforward part.
  We present a tailored online learning scheme of the Gaussian
  process, adapted to scanning quantum dot microscopy, that includes hyperparameter optimization
  during operation to enable fast and precise scanning of arbitrary surface
  structures.
  For the potential application in practice, the accompanying
  computational cost is reduced evaluating different sparse approximation
  approaches.
  The fully independent training conditional approximation, used on a
  reduced set of active training data, is found to be the most promising
  approach.
\end{abstract}

%%%%%%%%%%%%%%%%%%%%%%%%%%%%%%%%%%%%%%%%%%%%%%%%%%%%%%%%%%%%%%%%%%%%%%%%%%%%%%%%
% CONTENT

\input{Introduction.tex}
\input{SQDM.tex}
\input{2DOF.tex}

\input{SparseGP.tex}
\input{SimResults.tex}
\input{Conclusions.tex}

%%%%%%%%%%%%%%%%%%%%%%%%%%%%%%%%%%%%%%%%%%%%%%%%%%%%%%%%%%%%%%%%%%%%%%%%%%%%%%%% 

\addtolength{\textheight}{-12cm}   % This command serves to balance the column lengths
                                  % on the last page of the document manually. It shortens
                                  % the textheight of the last page by a suitable amount.
                                  % This command does not take effect until the next page
                                  % so it should come on the page before the last. Make
                                  % sure that you do not shorten the textheight too much.

%%%%%%%%%%%%%%%%%%%%%%%%%%%%%%%%%%%%%%%%%%%%%%%%%%%%%%%%%%%%%%%%%%%%%%%%%%%%%%%%
\bibliographystyle{IEEEtran}
\bibliography{IEEEabrv,Bibliography}

\end{document}

%% file: Introduction.tex
% Paper: SQDM Control with sparse GP
% Section 1: Introduction

\section{Introduction}
\label{sec:intro}

Research in physical chemistry and nanotechnology is driven by a wide
field of possible future applications such as, for instance, molecular
manipulation that enables the assembly of molecular machines
or nanoscopic electric circuits by single molecule placement
\cite{Moresco2015, Findeisen2016}.
This research requires a fundamental understanding of the basic building
blocks of matter, atoms and molecules.
Therein, imaging of nanostructures plays an important role.

\begin{figure}
	\centering
	\includegraphics[width=0.3\textwidth]{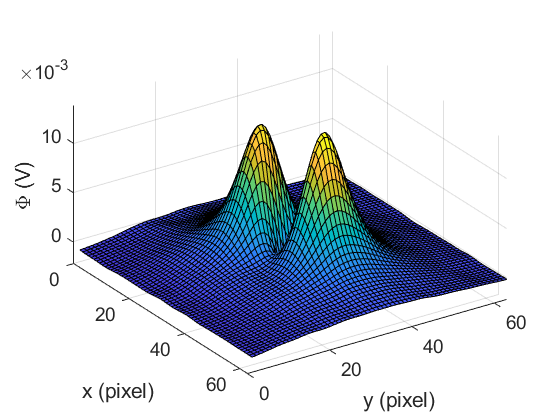}
	\caption{SQDM image of a single molecule \cite{Wagner2015}.}
	\label{f:intro}
\end{figure}

Visualizing surface nanostructures can be accomplished using scanning
probe microscopy techniques \cite{Salapaka2008}. 
One such recently developed technique is scanning quantum dot microscopy
(SQDM) \cite{Wagner2015, Wagner2019a} that probes the
electrostatic potential of a surface structure using a single
molecule as sensor. 
It provides qualitative and quantitative images of the \EPs of
nanostructures on surfaces with atomic resolution as shown in
Fig.~\ref{f:intro}.

Despite SQDM's large potential, widespread deployment in scientific
practice was initially hindered due to excessive scan times.
To accelerate the scanning process a tailored \DOF (2DOF) controller,
comprising a feedback and feedforward controller, was developed
\cite{Wagner2019a}.
The feedforward part is utilized to generate a prediction for upcoming
measurement points.
The feedback part corrects the discrepancies between the real and the
predicted measurement values.
So far, the feedforward part that is implemented in the experiment uses
the previously scanned line to generate a prediction for the next line.
In \cite{Maiworm2018} however, we could show in simulations that a
better prediction can be obtained if a Gaussian process (GP) is used to
generate the feedforward signal.

% Contribution
In this work we further develop the 2DOF controller with the GP towards
real-time applicability for SQDM.
To this end, two challenges in particular regarding the feedforward are
elementary.
First, so far, a priori knowledge of the involved
\emph{hyperparameters} of the GP was assumed to be available.
Second, the necessary GP computations are expensive for real-time
feasibility.
To account for these challenges the contributions of this work are
\begin{enumerate}
  \item online hyperparameter learning for SQDM such that no a priori
    knowledge is required,
  \item verification of real-time capability via preselection,
    comparison, and adaptation of different sparse GP approaches 
    for computational reduction, and
  \item verification of the selected approaches and their potential
    real-time feasibility in SQDM control.
\end{enumerate}

One of the main messages of this paper is that the fusion of control
with learning enables new and improved applications, such as SQDM.

The structure of the remainder of this paper is as follows.
We outline the working principle of SQDM in Section \ref{sec:sqdm}. In
Section \ref{sec:2dof}, we introduce the 2DOF control paradigm for SQDM
including a brief overview of GPs. In Section \ref{sec:sparsegp}, we
present the sparse GP approaches compared in this work. The simulation
results are shown in Section \ref{sec:simresults} and we draw
conclusions in Section \ref{sec:conclusions}.

%% file: SQDM.tex
% Paper: SQDM Control with sparse GP
% Section 2: Scanning Quantum Dot Microscopy

\section{Scanning Quantum Dot Microscopy}
\label{sec:sqdm}

\begin{figure}
	\centering
	\subfloat[][]{\includegraphics[width=0.3\textwidth]{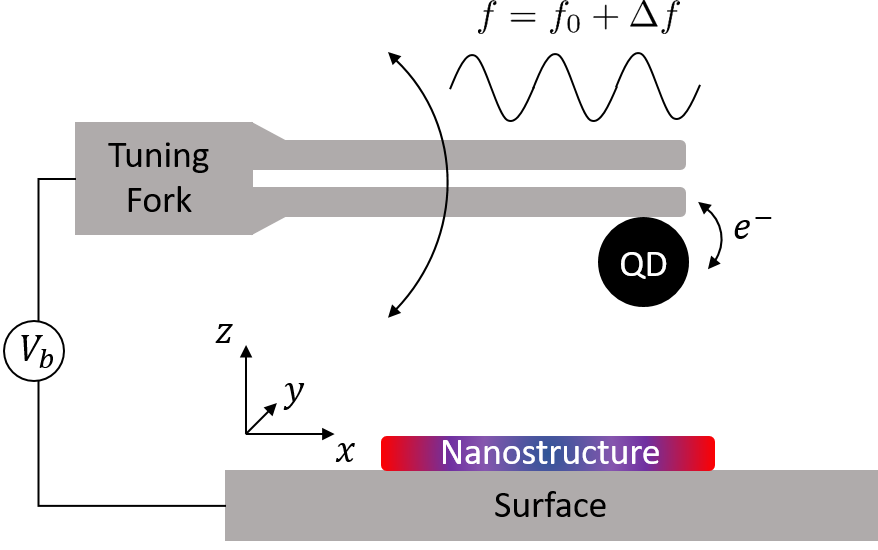}}
	\qquad
	\subfloat[][]{\includegraphics[width=0.14\textwidth]{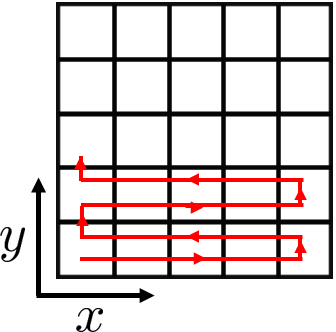}}
	\caption{\textbf{(a)}: Schematic set-up of \SQDM (adapted from \cite{Maiworm2018}). 
  \textbf{(b)}: Raster scan pattern. 
  The image is divided into pixels.
  Lines are indexed by $y$, columns by $x$. 
  The red arrows indicate scanning direction.
  }
	\label{f:sqdm}
\end{figure}

SQDM generates images of \EPs of nanostructures on surfaces with
nanometer resolution.
To this end it utilizes a sensor molecule\footnote{Currently, a single
perylene-3,4,9,10-tetracarboxylic dianhydride (PTCDA) molecule is used.}
denoted as quantum dot (QD)
\cite{Wagner2015, Wagner2019a}, which is bonded to the 
tuning fork of a frequency modulated non-contact atomic force
microscope (NC-AFM).
Between the sample and the tip a bias voltage source \Vb is connected
(Fig.~\ref{f:sqdm}a) that generates an \EP, which is superimposed to that
of the sample.

\begin{figure}[b]
	\centering
	\includegraphics[width=\linewidth]{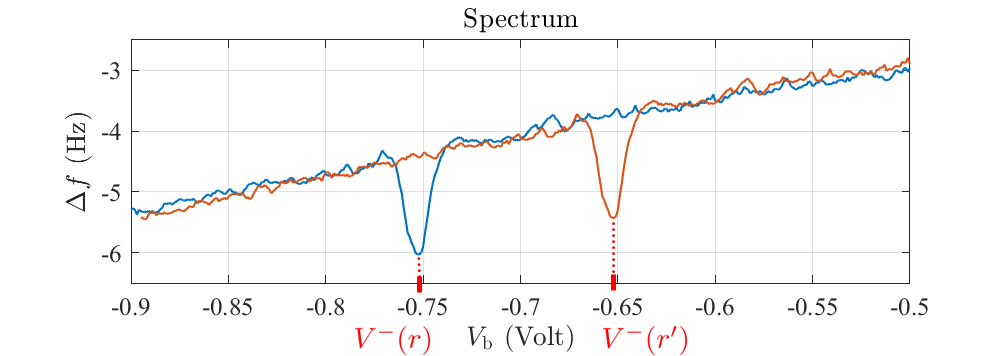}
  \caption{Exemplary cutout of the spectrum $\Df(\Vb)$ illustrating the
  movement of the negative dip at two microscope tip positions $r$ and $r'$.
  Each plotted spectrum belongs to one specific tip position.
  For an illustration of the complete spectrum with both dips, see
  \cite{Maiworm2018}.
  }
	\label{f:spectrum}
\end{figure}

While performing a raster scan pattern (Fig.~\ref{f:sqdm}b), \Vb
is varied.
If the effective potential at the QD reaches specific
thresholds, the QD is charged or discharged via electron tunneling
between the microscope tip and the QD.
These charging events lead to abrupt changes in the tuning fork's
oscillation frequency, sensed by the NC-AFM, and appear in the
\emph{spectrum} $\Df(\Vb)$ as features, called \emph{dips}
(Fig.~\ref{f:spectrum}).
The two occurring dips, one at negative voltages and one at positive
voltages, are characterized by their respective position in the
spectrum, indicated by their minima \Vneg and \Vpos, or \Vmp for short.
The $\Vmp(r)$ values change with the tip position $r = (x,y,z)$ and are
the main measurements of SQDM\footnote{Note that images are usually
generated at a constant height $z$. Therefore we omit the dependence 
on $z$ in the following.}, which are then used in a post-processing
step to determine the \EP of the sample \cite{Wagner2019a, Wagner2019b}.
Note that the sample has to be scanned twice because the $\Vmp(r)$ data
can be obtained only separately.

The challenge of SQDM imaging lies in the a priori unknown and changing
voltage values $\Vmp(r)$ during scanning, which depend both on the
sample's topography, as well as its electric properties.
Originally, the $\Vmp(r)$ values were determined by varying \Vb in a broad
interval at each image pixel, which results in excessive scan times.
This problem can be circumvented by employing a tailored \DOF control
approach that continuously adapts \Vb to directly track the dips
\cite{Maiworm2018}.

%% file: 2DOF.tex
% Paper: SQDM Control with sparse GP
% Section 3: 2DOF Control of SQDM including GPs

\section{Two-Degree-Of-Freedom Control of SQDM}
\label{sec:2dof}

In this section, we first outline the 2DOF control paradigm
and its application to SQDM. 
Secondly, we give a brief overview of Gaussian processes including
hyperparameter optimization.

\subsection{2DOF Control Paradigm}
	
\begin{figure}
	\centering
	\includegraphics[width=0.45\textwidth]{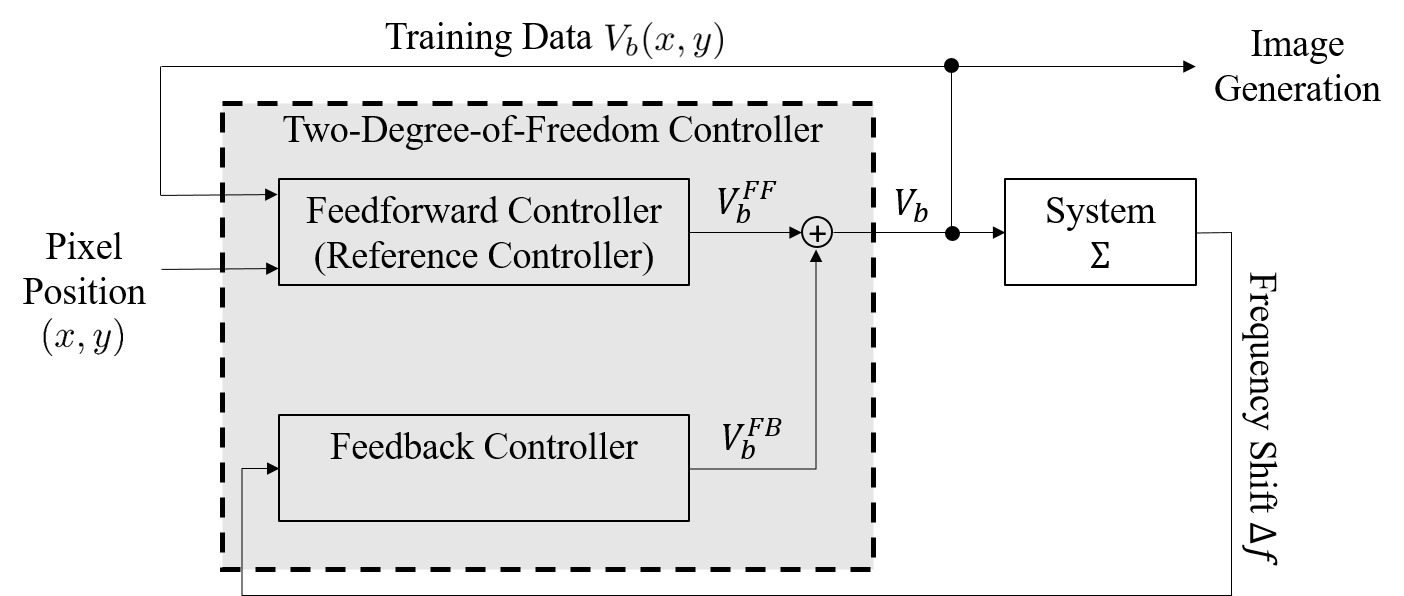}
  \caption{ Diagram of the SQDM two-degree-of-freedom controller. 
  The system $\Sigma$ represents the NC-AFM.
  The frequency shift \Df of the NC-AFM is the feedback signal
  for the feedback controller, c.f. \cite{Maiworm2018}.
  }
	\label{f:2dof}
\end{figure}

The \DOF controller as presented in \cite{Maiworm2018} consists of a
feedback and a feedforward part (Fig.~\ref{f:2dof}).
The feedback controller is an \ESC \cite{Krstic2000b} that allows to
directly track the dips and with that the $\Vmp(r)$ values.
It continuously adjusts \Vb such that the respective dip is minimized
\cite{Maiworm2018}, \ie
\begin{equation}
	\begin{aligned}
  		\Vneg(r) &= \argmin_{\Vb < 0} \Df(\Vb,r) \\
  		\Vpos(r) &= \argmin_{\Vb > 0} \Df(\Vb,r) \ .
	\end{aligned}
	\label{e:Vpm}
\end{equation}

Tracking of the dips with the \ESC works only as long as the
current \Vb value is within the respective dip.
If one of the dips changes its position faster (see Fig.~\ref{f:spectrum})
than the feedback controller is able to follow, then \Vb leaves the dip
and scanning has to be aborted.
To prevent this, the feedforward part, which is based on a Gaussian
process, generates a prediction $\Vb^\text{FF}$ of the respective
$\Vmp(r)$ evolution for the next line, which is added to the \Vb output
of the feedback controller $\Vb^\text{FB}$.
Thus, the feedforward signal has to be as accurate as possible and is
critical for correct operation.
The remainder of this work will be focused therefore on using Gaussian
process based learning to obtain the feedforward signal $\Vb^\text{FF}$ for the next line, based on the data of previous lines.

\subsection{Gaussian Process Regression}
	
A Gaussian process $f(\xi) \sim \GP (m(\xi) , k(\xi, \xi'))$ can be
used to model a function $f: \mathbb{R}^t \rightarrow \mathbb{R}^s, \xi
\mapsto f(\xi)$.
In this work it will be used to model the feedforward signal with the
tip position $\xi = r = (x,y)$ as input and the respective output $f(\xi) =
\Vmp$. 
Formally, a Gaussian processes is defined as \textit{a collection of random variables, any
finite number of which have consistent joint Gaussian distributions}
\cite{Candela2007}. That means, loosely speaking, that one might think of
a Gaussian process as being an infinite-dimensional, multivariate normal
distribution where the function values $f(\xi)$ are random values
evaluated at the inputs $\xi$.
A GP is fully defined by its mean function $m(\xi) = E \lbrack f(\xi)
\rbrack$ ($E [ \cdot ]$ denotes the expected value) and covariance function
$k(\xi , \xi') = \text{cov} \lbrack f(\xi) , f(\xi') \rbrack = E
\lbrack (f(\xi) - m(\xi)) (f(\xi') - m(\xi')) \rbrack$, which both
depend on a set of so-called hyperparameters $\theta$
\cite{Rasmussen2006}, \cite{Kocijan2016}. 

The objective is to learn the function $f$ and compute at $\nt$ test inputs $\xi_*^{(j)},~ j = 1,...,\nt$ the corresponding, predicted test targets stored as a vector $f_* \in \R^{\nt}$. 
To this end, noisy training observations (training targets) 
$\y = f(\Xi) + \varepsilon$ are required, where $\Xi \in \R^{d \times n}, \Xi = \lbrack \xi^{(1)}, ... ,\xi^{(n)} \rbrack$ 
is the matrix storing $n$ $d$-dimensional training inputs, $f(\Xi) \in \R^n$ is the vector of the corresponding, 
noise-free function values and $\varepsilon \sim \N (0, \Sn \I)$, 
% \footnote{$\N (\cdot , \cdot \cdot)$ is a Gaussian probability distribution with mean $( \cdot )$ and variance $( \cdot \cdot )$.}
with $\I$ 
being the identity matrix, models white Gaussian/normally distributed
noise with variance \Sn. 
The training and test data points are assumed to be independently and
identically distributed.

\begin{figure}
	\centering
	\subfloat[][]{\includegraphics[width=0.25\textwidth]{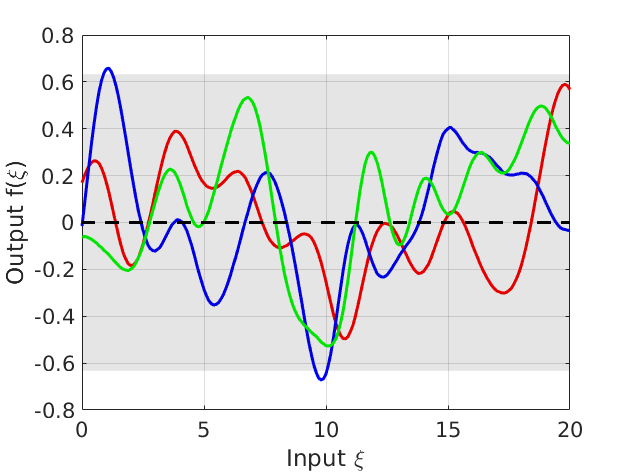}}
	\subfloat[][]{\includegraphics[width=0.25\textwidth]{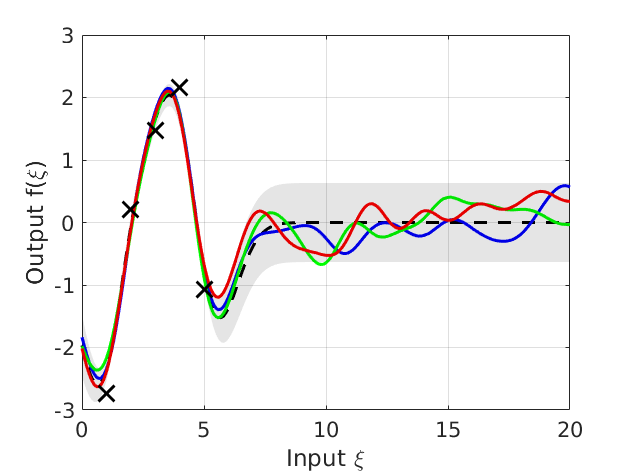}}
	\caption{Illustration of GP inference.
  Functions drawn from a zero mean GP with isotropic squared exponential covariance function ($\sigma_f = 0.1$, $\ell = 1.0$). The grey-shaded area indicates the 95 \% confidence interval. The dashed line shows the mean function. \textbf{(a)}: Random functions drawn from the prior distribution. \textbf{(b)}: Random functions drawn from the posterior distribution (\ref{e:gpposterior}) with noise variance $\Sn = 0.01$. Black crosses indicate training data.}
	\label{f:gp}
\end{figure}

To learn the function $f$, all prior functions (Fig.~\ref{f:gp}a) that
cannot explain the measured training data set $\{ \Xi, \y \}$ have to be
rejected (Fig.~\ref{f:gp}b).
This \emph{inference} step is mathematically done by conditioning
the prior on the training data (\cf \cite{Rasmussen2006}) to arrive at
the posterior distribution $f_* \mid \Xi, \y , \Xi_* \sim \N (\bar{f}_* ,
\hat{\sigma}^2)$ with
\begin{subequations}
	\begin{align}
		& \bar{f}_* \! = \! m_{\Xi_*} \! + \! K(\Xi_*,\Xi) K_{\y} \inv K(\Xi,\Xi_*) (\y \!
      - \! m_{\Xi}) \label{e:postMean} \\
		& \hat{\sigma}^2 \! = \! K_* \! - \! K(\Xi_*,\Xi) K_{\y} \inv K(\Xi,\Xi_*) \ ,
	\end{align}
	\label{e:gpposterior}
\end{subequations}
where $\bar{f}_* \in \R^{\nt}$ is the posterior mean prediction and $\hat{\sigma}^2 \in \R^{\nt \times \nt}$ is the predictive covariance matrix. 
Furthermore, $\Xi_* \in \R^{d \times \nt}, \Xi_* = \lbrack \xi_*^{(1)}, ... , \xi_*^{(\nt)} \rbrack$ is the matrix storing $\nt$ $d$-dimensional test inputs, $K(\cdot , \cdot )$ are covariance matrices, $K_{\y} = K(\Xi,\Xi) + \Sn \I$, $K_* = K(\Xi_*,\Xi_*)$, $m_{\Xi_*} = m(\Xi_*)$ and $m_{\Xi} = m(\Xi)$.

Thus, the GP prediction used for the SQDM feedforward signal is
computed by \eqref{e:postMean}, \ie based on previous measurement pairs
$\big( (x,y), \Vmp \big)$ the GP builds a model for $\Vmp(x,y)$ and
generates a prediction for the next line $y+1$.

\subsection{Hyperparameter Adaptation}

The GP mean and covariance function depend on a set of
hyperparameters $\theta$ and hence, an appropriate GP prediction
requires suitable hyperparameters.
The optimal hyperparameters $\theta^*$ for a certain training data set
$\{\Xi,\y\}$ can be obtained by maximizing the logarithmic likelihood 
\begin{align*}
  \text{log} \big( p(\y \mid \Xi, \theta) \big) = - \frac{1}{2} {\y_0}^T K_{\y}
    \inv \y_0 \! - \! \frac{1}{2} \ln |K_{\y}| \! - \! \frac{n}{2} \ln (2\pi) ~ ,
\end{align*}
where $p( \cdot )$ is the probability density function, $| \cdot |$ denotes the determinant and $\y_0 = \y - m_{\Xi}$ is a
standardization \cite{Rasmussen2006}.
The logarithmic likelihood describes how well the training data is
explained by the underlying model.
The optimal hyperparameters are then
\begin{align}
  \theta^* = \arg \max_{\theta} \left\{ \text{log}\big( p(\y \mid  \Xi,
             \theta ) \big) \right\} \ .
	\label{e:opt}
\end{align}

In SQDM, the training data is a continuously incoming stream of
measurements.
Therefore, hyperparameter optimization has to be performed online such
that the GP model is capable of adequately describing the evolution of
$\Vmp(r)$ in the current region of operation without a priori knowledge.
% comp. cost
This is computationally challenging because the computation time
required for $K_{\y}^{-1} \in \R^{n \times n}$ scales in general with $\CC
(n^3)$ \cite{Rasmussen2006}.

%% file: SparseGP.tex
% Paper: SQDM Control with sparse GP
% Section 4: Efficient Implementation of Gaussian Processes

\section{Efficient Implementation of Gaussian Processes}
\label{sec:sparsegp}

\begin{figure}[t]
	\centering
	\includegraphics[width=0.45\textwidth]{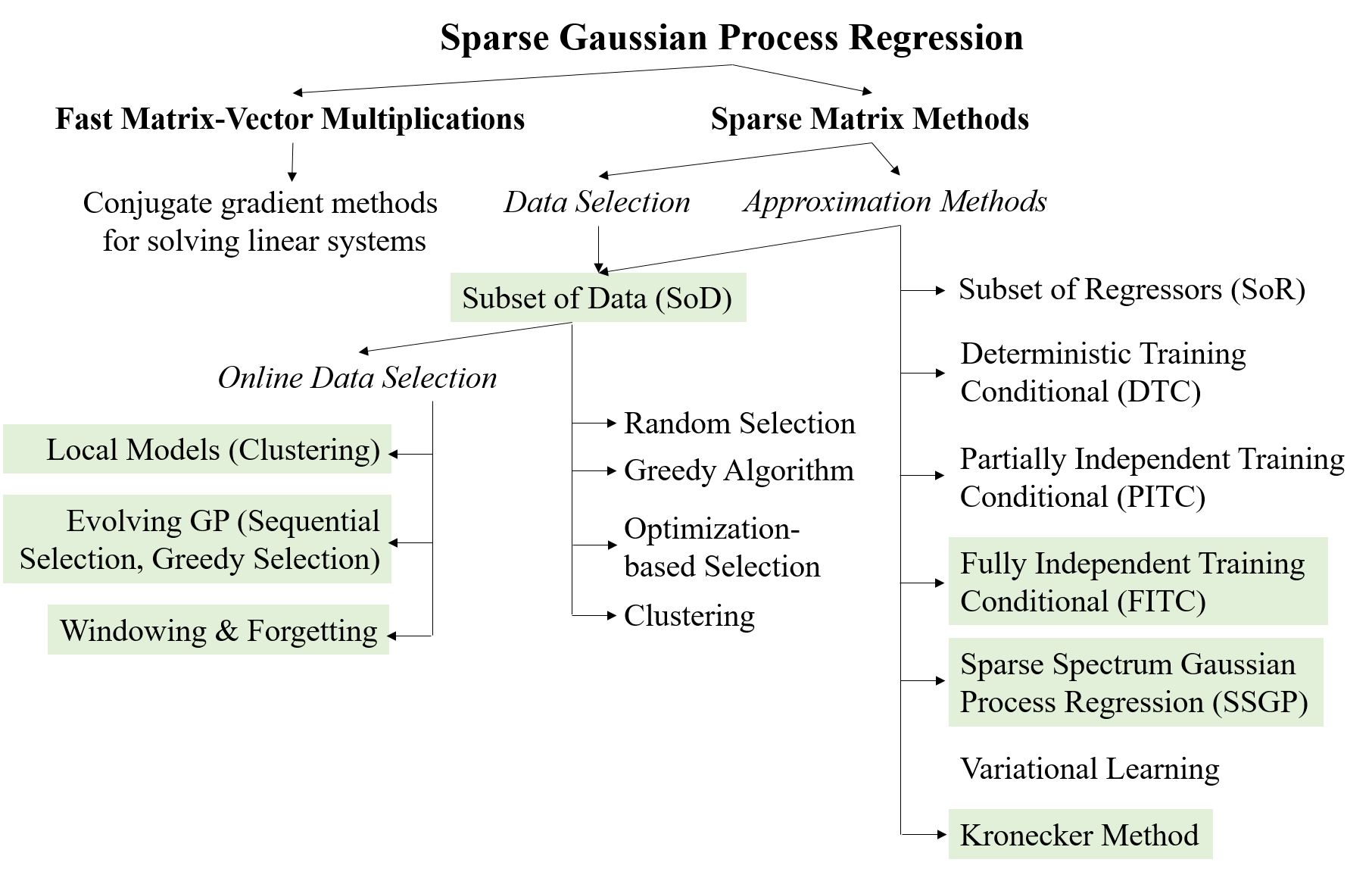}
	\caption{Overview of approaches for a sparse implementation of Gaussian processes. The ones selected for implementation and testing are highlighted.}
	\label{f:overview}
\end{figure}

Computing and adjusting the GP online is crucial for use in SQDM.
To reduce the computational load, especially for hyperparameter
optimization as mentioned in the previous section, we have identified
suitable approaches in a literature review.
Fig.~\ref{f:overview} provides an overview of the identified approaches,
while more detailed presentations can be found in \cite{Kocijan2016} (overview), \cite{Gredilla2010} (SSGP), \cite{Candela2007} (overview), \cite{Rasmussen2006} (overview), \cite{Sigaud2010} (clustering), \cite{Smola2000} (SoR), \cite{Seeger2003} (DTC), \cite{Snelson2006} (FITC), \cite{Titsias2009} (variational learning) and \cite{Wilson2015} (structure exploiting approaches, especially Kronecker method).
Four of the approaches have been selected
as the most promising for application in SQDM control and are therefore outlined in the following.

	\subsection{Subset of Data}

\begin{figure}
	\centering
	\includegraphics[width=0.3\textwidth]{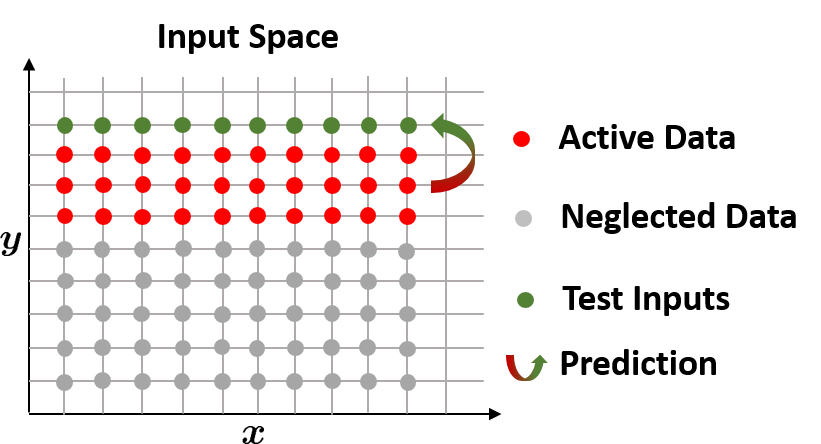}
	\caption{In the SoD approach, the prediction is based only on a subset of the available training data. Here, the sliding window approach for data selection is shown.}
	\label{f:sod}
\end{figure}

Instead of using the entire available training data, \ie the training set $\mathcal{X} = \lbrace \xi^{(j)} , \y_j \rbrace _{j = 1}^n$,
only a smaller subset of training data, the active set $\mathcal{I} = \lbrace \xi^{(h)}, \y_h \rbrace_{h = 1}^{\nI}$ of
size $\nI \ll n$, is used in the SoD approach for GP modeling (Fig.~\ref{f:sod}). 
This reduces the computational complexity from $\CC (n^3)$ to $\CC (\nI^3)$ \cite{Candela2007}, \cite{Chalupka2013}.
Within this work, we use three different approaches outlined in the following for building the active set $\mathcal{I}$.

\textit{Sliding Window}: The active set consists of the $\nI$ most recent data points \cite{Vaerenbergh2006}.

\textit{Evolving GP}: A new data point is included in $\mathcal{I}$ only if it provides new information to the current GP model, \ie if the prediction error or the predictive variance exceed preset thresholds. 
Any time a new data point is included in $\mathcal{I}$, the point with the lowest information gain is removed \cite{Kocijan2016}. 
For simplification, we just remove the oldest data point.

\textit{Clustering}: The training data is clustered using the \emph{k-means} algorithm (\cf \cite{Barbakh2009}) with the covariance function as similarity measure.
These clusters are then the active sets for distinct GP models \cite{Sigaud2010}.

	\subsection{Kronecker Method}
	
\begin{figure}[b]
	\centering
	\includegraphics[width=0.3\textwidth]{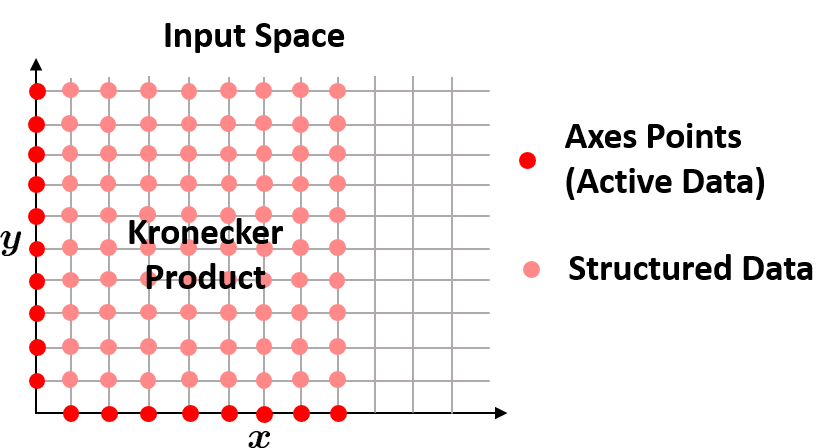}
	\caption{In the Kronecker method, the grid structure of the training data is exploited by using only the axes points as active points. The entire grid data is computed as Kronecker product.}
	\label{f:kron}
\end{figure}

The Kronecker method can only be applied if the inputs are points of a $D$-dimensional Cartesian grid and the 
covariance function $k$ has a product structure across grid dimensions. 
Exploiting the covariance function's 
structure, it can be rewritten as $k(\xi,\xi') = \prod_{d=1}^D k (\xi_d, \xi'_d)$. 
Then, the covariance 
matrix $K \in \R^{\nI \times \nI}$, where $\nI$ is the number of grid points, can be computed using the Kronecker 
product $K = K_1 \otimes ... \otimes K_D$, where $K_d ~,~ d = 1, ... ,D$ is the covariance matrix of the points on 
the grid's $d^{th}$ axis (Fig.~\ref{f:kron}). 
As a result of their smaller size, the Kronecker factors $K_d$ can 
be quickly eigendecomposed. From their eigendecompositions $K_d = Q_d V_d Q_d^T$, the eigendecomposition of $K$ 
can be efficiently computed as $K = Q V Q^T$ with $Q = Q_1 \otimes ... \otimes Q_D$ and $ V = V_1 \otimes ... \otimes V_D$. 
Therewith, computing the inverse $\lbrack K + \Sn \I \rbrack \inv$ becomes trivial and the 
computationally complexity reduces to $\CC \left( D \nI^{( 1 + \frac{1}{D} ) } \right)$ \cite{Wilson2015}.

	\subsection{Fully Independent Training Conditional}

\begin{figure}
	\centering
	\includegraphics[width=0.3\textwidth]{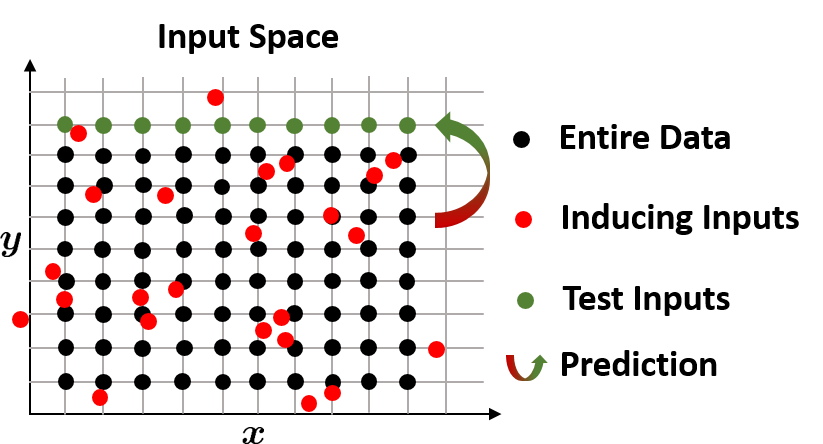}
	\caption{In the FITC approach, a small pseudo data set is used to explain the entire training data.}
	\label{f:fitc}
\end{figure}

The Fully Independent Training Conditional (FITC) approximation is based
on the idea to use a pseudo data set of $\nI \ll n$ inducing inputs stored
in the matrix $I \in \R^{d \times \nI}$ and the corresponding observations
$f_I \in \R^{\nI}$ to explain the training and test data (Fig.~\ref{f:fitc}). 
There are two critical assumptions underlying the FITC
approximation (\cf \cite{Candela2007}), which we also adopt in this work. 
The first is that one assumes the training and test observations $\y$ and $f_*$ to be \textit{conditionally independent given} $f_I$, \ie that
\begin{equation*}
	\begin{aligned}
		p(f_*,\y) & = \int p(f_*,\y ,f_I) \diff f_I \\
		& = \int p(f_* \mid f_I) p(\y \mid f_I) p(f_I) \diff f_I ~
	\end{aligned}
\end{equation*}
holds. This means, loosely speaking, that the test and training variables are only connected indirectly via the 
inducing variables that \textit{induce} their dependencies \cite{Candela2007}. 
One might think of projecting the entire training variables onto the inducing variables and compute the test 
variables again as a projection of the inducing variables, such that the prediction is based on an approximate 
representation of the entire training data. 

The second assumption affects the relationship between the training and the inducing variables described by the 
conditional $p(\y \mid f_I)$. 
A fully independent conditional is assumed, \ie the training variables are only self-dependent and thus, the 
covariance matrix has a diagonal structure. The approximate conditional $q(\y \mid f_I)$ is then given by
\begin{equation*}
	q(\y \mid f_I) \! = \! \N \left( m_{\Xi} \! + \! K(\Xi,I) K_I \inv f_{I,0} ~,~ \text{diag} \lbrack K_{\y} \! - \! Q_{Xi} \rbrack \right) ,
\end{equation*}
where $Q_{\Xi} = K(\Xi,I) K_I \inv K(I,\Xi)$, $K_I = K(I,I)$ and $f_{I,0} = f_I - m(I)$ \cite{Candela2007}, 
\cite{Snelson2006}.
Finally, the predictive distribution using FITC is
\begin{equation*}
	\begin{aligned}
		q_{\text{FITC}}(f_* \mid \y) & = \N \left( \bar{f}_{*,\text{FITC}}, \hat{\sigma}^2_{\text{FITC}} \right) \\
		\bar{f}_{*,\text{FITC}} & = m_{\Xi_*} \! + \! K(\Xi_*,I) \Sigma K(I,\Xi) \Lambda \inv \y_0\\
		\hat{\sigma}_{\text{FITC}}^2 & = K_* - Q_* + K(\Xi_*,I) \Sigma K(I,\Xi_*) ~, \\
	\end{aligned}
\end{equation*}
where $\Sigma = \left \lbrack K_I + K(I,\Xi) \Lambda \inv K(\Xi,I) \right \rbrack \inv$, $ \Lambda = \text{diag} \lbrack K_{\y} - Q(\Xi,\Xi) \rbrack$ and $Q_* = K(\Xi_*,I) K_I \inv K(I,\Xi_*)$ \cite{Candela2007}. 

The inducing inputs can be considered as additional covariance hyperparameters and thus, they can be found 
optimization-based by solving (\ref{e:opt}) \cite{Candela2007}, \cite{Snelson2006}. \\
The computational complexity of the FITC approximation is $\CC (n \nI^2)$ \cite{Candela2007}.

	\subsection{Sparse Spectrum Gaussian Process Regression}
	
\begin{figure}
	\centering
	\includegraphics[width=0.23\textwidth]{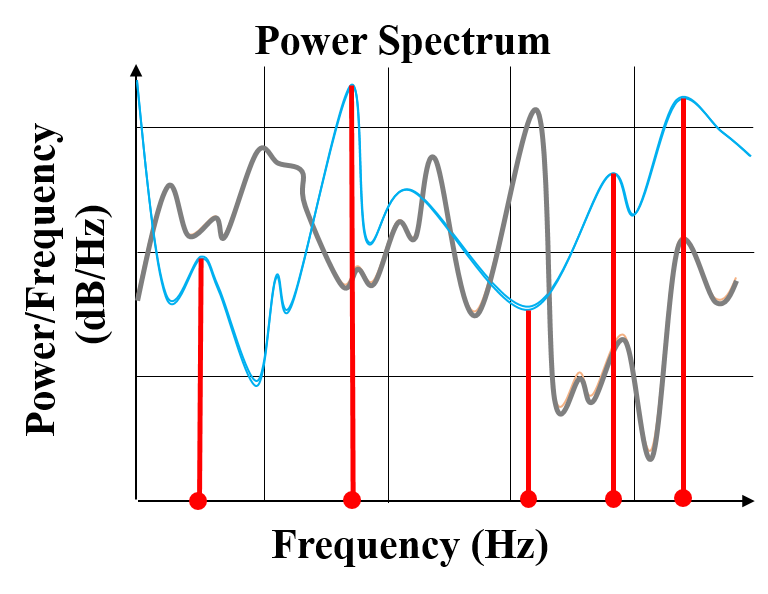}
	\caption{In SSGPR, the power spectrum of the GP is learned and sparsely represented by a finite set of Dirac pulses with finite amplitude.}
	\label{f:ssgp}
\end{figure}

In Sparse Spectrum Gaussian Process Regression (SSGPR), the GP is Fourier transformed into the frequency domain in 
order to decompose it in a (infinite) set of oscillations, each with a certain frequency. 
Thus, SSGPR works with power spectra rather than with random functions directly. 
A power spectrum of the GP describes how strong each frequency contributes to it (Fig.~\ref{f:ssgp}). 

The basic idea of SSGPR is now to approximate the GP's power spectrum by a finite set of Dirac pulses with finite 
amplitude to sparsify its representation \cite{Gredilla2010}. 
We therefore consider a trigonometric Bayesian regression model given by
\begin{equation}
	f(\xi) = \sum_{r = 1}^{\nI} a_r \cos (2\pi s_r^T \xi) + b_r \sin (2 \pi s_r^T \xi)
\end{equation}
with $a_r \sim \N \left( 0, \frac{\sigma_f^2}{\nI} \right)$ and $b_r \sim \N \left( 0, \frac{\sigma_f^2}{\nI} \right)$, where $a_r$ and $b_r$ are the amplitudes of the basis functions, $s_r$ is a vector of spectral frequencies, 
$\nI$ is the number of basis functions and $\sigma_f$ is a covariance hyperparameter (\cf Sec. \ref{sec:simresults}). 
Under these conditions, the distribution over functions is a zero-mean Gaussian with the stationary covariance 
function \cite{Gredilla2010}
\begin{equation}
	k(\xi , \xi') = \frac{\sigma_f^2}{\nI} \sum_{r = 1}^{\nI} \cos (2 \pi s_r^T (\xi - \xi')) ~~ .
\end{equation}
The spectral frequencies are therefore additional covariance hyperparameters describing the positions of the Dirac 
pulses. 
In consequence, the spectral frequencies can be found optimization-based by maximizing the logarithmic likelihood 
(\cf (\ref{e:opt})) \cite{Gredilla2010}. 
That means that the spectral frequencies are selected given the training data and thus, we learn the power 
spectrum of the GP instead of its mean function. 
Although the standard representation of the predictive distribution and the logarithmic likelihood as given in 
Section \ref{sec:2dof} can be used in SSGPR, a more efficient one is provided in \cite{Gredilla2010}. 

The computational complexity of SSGPR is $\CC (n \nI^2)$ \cite{Gredilla2010}.

%% file: SimResults.tex
% Paper: SQDM Control with sparse GP
% Section 3: 2DOF Control of SQDM including GPs

\section{Implementation and Simulation Results}
\label{sec:simresults}

In this section, we first discuss the implementation of the online hyperparameter adaptation. 
Secondly, we show the simulation results for different reference images when using the sparse GP approaches 
presented in Section \ref{sec:sparsegp}. 
Thirdly, we will investigate the computation times and practical applicability of the approaches.

	\subsection{Preliminaries}
	
We use a GP with constant mean function $m(\xi) = c$ with $c \in \R$. 
We tested several covariance functions of which the squared
exponential (SE) covariance function $k(\xi_p , \xi_q) = \sigma_f^2 \exp
\left( -\frac{1}{2} (\xi_p - \xi_q)^T M (\xi_p - \xi_q) \right)$ worked
best\footnote{
An important property of SE covariance function is that it
yields smooth functions.
Since \EPs are always smooth, the SE covariance function is an
appropriate choice.
}, with $M = \ell^{-2} \I, \ell \in \R_{>0}$ for the isometric form
(used for FITC and SSGPR) and $M = \text{diag}(\ell)^{-2}, \ell \in
\R^D_{>0}$ for the form with automatic relevance determination
(used for SoD and the Kronecker method) and $\sigma_f \in \R_{>0}$
\cite{Rasmussen2006}.
The algorithms used for implementation are geared towards the ones of the GPML toolbox \cite{GPML}.

During first evaluations we have found that FITC and SSGP work best on a
reduced active training data set of the last five scanned lines
and with one third of the number of training data as number of inducing
inputs and spectral points respectively. For SoD and the Kronecker
method, we use active sets of the size of two lines.

	\subsection{Online Hyperparameter Adaptation} 

In order to make SQDM universally applicable, the controller has to have 
the ability of adapting to a specific experimental setting, e.g., arbitrary 
surface samples.
In particular, the hyperparameters have to be adapted during scanning when no or only 
little prior knowledge on the scanned surface structure is available. 
Besides the hyperparameters' dependency on the surface structure, they further 
depend on the physical properties of the quantum dot that is used for scanning. 
In addition, if the GP is sparsified using an inducing point method such as FITC, 
the number of inducing 
inputs should be chosen according to the amount of training data and thus, the number of hyperparameters depends 
in such case also on the size of the chosen raster pattern. 

Hyperparameter adaptation is computed via (\ref{e:opt}) and implemented according to the following scheme. 
In a first step, the partial derivatives of the logarithmic likelihood w.r.t. the hyperparameters are 
computed. 
Additional prior knowledge on the hyperparameters can be included in the second step by correcting the 
derivatives due to the respective prior distributions. In a third step, the partial derivatives are 
then used in a conjugate gradient (CG) method with search direction computed according to 
Polak-Ribi\`{e}re (\cf \cite{Chang2015}) to compute an estimate of the optimal hyperparameters. 
After the CG iteration has finished, the current estimate for the optimal hyperparameters is returned to the SQDM model (\cf \cite{Maiworm2018}) and used for prediction.

As hyperparameter learning is computationally expensive, we have to approximate the GP using the 
approaches presented in Sec. \ref{sec:sparsegp} in the next step to
reduce computation to allow for online implementation.

	\subsection{Comparison of Sparse GP Approaches}
	
\begin{figure}
	\centering
	\begin{tabular}{c}
		\includegraphics[width=0.3\textwidth]{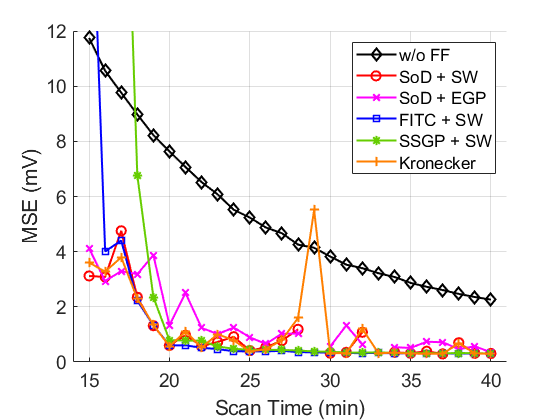} \\
		\footnotesize (a) \\
		\includegraphics[width=0.3\textwidth]{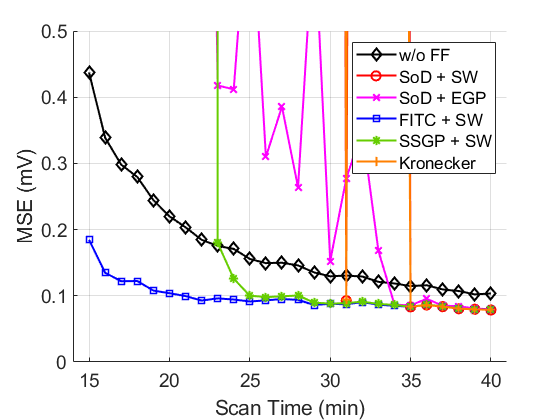} \\
		\footnotesize (b)
	\end{tabular}
	\caption{MSE for scanning R1 at different scan times. w/o FF: ESC only. SoD: Subset of Data. Kronecker: Kronecker method. FITC: Fully Independent Training Conditional. SSGP: Sparse Spectrum Gaussian Process. SW: Sliding Window. EGP: Evolving GP. \textbf{(a)}: $V^+$ scan. Missing data points indicate an interruption of the simulation due to an error. \textbf{(b)}: $V^-$ scan.}
	\label{f:R1scan}
\end{figure}

\vspace{-5pt}	
	
In order to compare the different sparse GP approaches, we simulate
first the scans of the reference image of Fig.~\ref{f:intro}, denoted as R1
for different scan times.  
For each scan, we quantify the image quality using the mean-squared error (MSE) (\cf \cite{Avcbas2002}, 
\cite{Sheikh2006}). 
For the sake of a better assessment of the results, we also show the MSE
when using feedback control only.

We start with feedback control only.
The MSE decreases monotonically with increasing scan time
(Fig.~\ref{f:R1scan}). 
As the scan time increases the ESC has more time to converge to the true
\Vmp values, which explains this observation.

We continue with the SoD approach and the \Vpos scan (Fig.~\ref{f:R1scan}a).
The MSE shows no clear and consistent trend as a function of the scan time but increases and decreases by leaps 
and bounds.
For the \Vneg scan (Fig.~\ref{f:R1scan}b), long scan times are required
for obtaining usable results\footnote{
  Performance differences of the 2DOF controller between the $V^{\mp}$
  scans are due to $\Vpos(r)$ and $\Vneg(r)$ being different functions
  (\cf \eqref{e:Vpm}) that are learned by the GP. 
  In the case of R1 the \Vneg map is more complex than the \Vpos
  map, \cf \cite{Maiworm2018} for an illustration.}.  
These are the same for either using the sliding window or the evolving GP approach for data selection.
SoD does not work in combination with clustering for building the active set.
As the input data comprises points on a Cartesian grid, there are no clusters that can be assigned appropriately.

Using the Kronecker method, the results for both the \Vneg and the
\Vpos scan are almost the same compared to the SoD results (Fig.
\ref{f:R1scan}) because in both approaches, in principle, the same
computations are performed. 
The small differences can be explained by the fact that each grid
dimension has a separate hyperparameter $\sigma_f$.

Contrary to the results obtained so far, a clear and consistent trend of decreasing
MSE values as well as small MSE values for low scan times is observed when using
FITC and SSGPR (Fig.~\ref{f:R1scan}).
FITC enables the lowest scan times while maintaining the highest outcome quality. 

\begin{figure}[b]
	\centering
	\subfloat[][]{\includegraphics[width=0.25\textwidth]{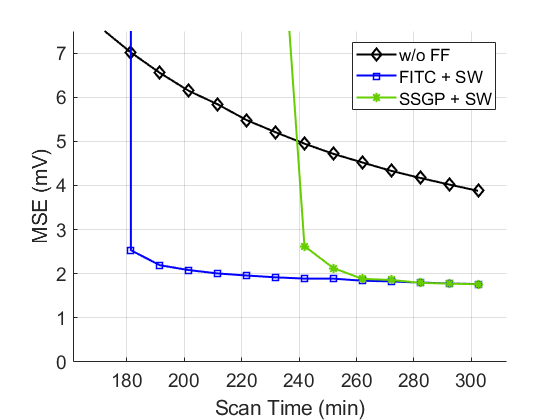}}
	\subfloat[][]{\includegraphics[width=0.25\textwidth]{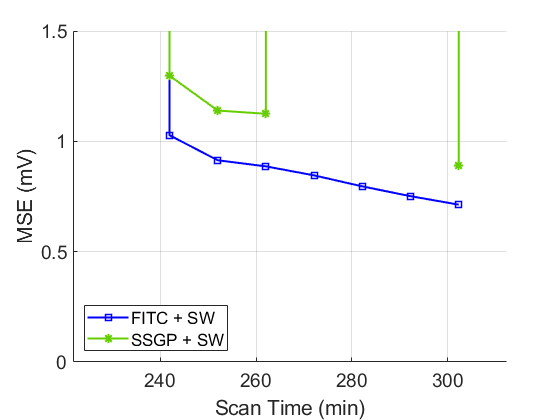}}
	\caption{MSE for scanning R2 at different scan times. w/o FF: ESC only. FITC: Fully Independent Training Conditional. SSGP: Sparse Spectrum Gaussian Process. SW: Sliding Window. \textbf{(a)}: \Vpos scan. \textbf{(b)}: \Vneg scan.}
	\label{f:R2scan}
\end{figure}

~\\
Now we simulate the scan of a second reference image, denoted as R2 
(\cf \cite{Wagner2019a}), which shows a 
different surface structure and is about ten times larger than
R1\footnote{This is the reason for higher scan times of R2 when compared
to R1.}.

Regarding the simulation results, both the SoD and the Kronecker 
approach fail at tracking the reference for the tested scan times.
Hence, we omit the corresponding data in Fig.~\ref{f:R2scan}.

For the remaining approaches we observe for the \Vpos scan (Fig.~\ref{f:R2scan}a) 
the same behavior as before for R1.
For the \Vneg scan (Fig.~\ref{f:R2scan}b) however, different observations are made.
The ESC without feedforward control fails to track the
reference for the tested scan times, which is why we omit showing the corresponding data in Fig.~\ref{f:R2scan}b.
We further observe that SSGPR has several problems at certain scan times, resulting unreasonably high errors.

~\\
Based on the presented simulation results, we can conclude that the FITC
has the largest potential for deployment in SQDM control.
	
	\subsection{Computation Times and Applicability}
	
\begin{figure}
	\includegraphics[width=0.485\textwidth]{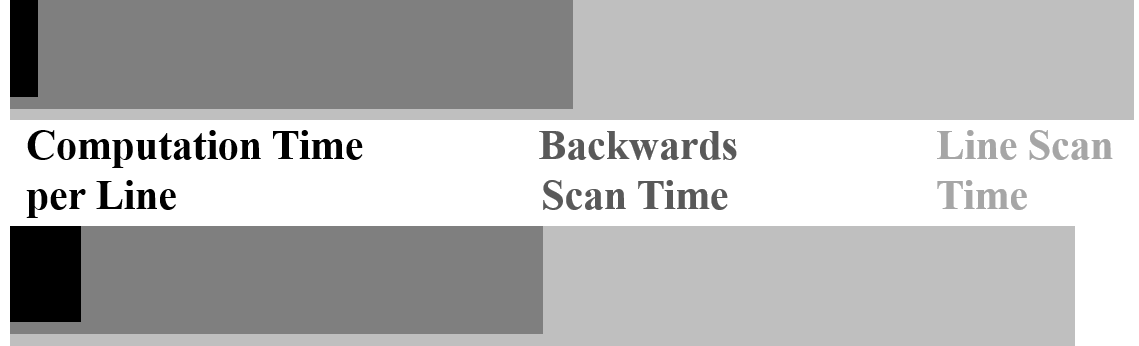}
	\caption{Average computation time per line in relation to the available time per line (backwards scan time) and the total line scan time. \textbf{Top}: \Vpos scan of R1 at \unit[20]{min} total image scan time. \textbf{Bottom}: \Vpos scan of R2 at \unit[180]{min} total image scan time.}
	\label{f:comptime}
\end{figure}

\vspace{-5pt}

As a last step, we investigate the computation times of FITC as
it yielded the best results in terms of performance.
Exemplarily, in the following we show the results for the \Vpos scans of
R1 and R2 (Fig.~\ref{f:comptime}), each with the minimal possible scan
time (R1: \unit[20]{min}, R2: \unit[180]{min}, \cf
Fig.~\ref{f:R1scan},~\ref{f:R2scan}).

We start with the $63 \times 63$ pixels image R1.
At a total scan time of \unit[20]{min}, the scan time per line is 
\unit[19.0]{s}. 
As in the experiments and due to the microscope's software,
each line is scanned back and forth.
Thus, half of the scan time is for scanning forward (data collection)
and half, \ie \unit[9.5]{s}, is for scanning backward.
The backwards scan time is the time per line that is
available for computing hyperparameter learning and inference. 
For R1, we find this computation time using FITC
to be in total \unit[28.1]{s}. This equals an average computation time per
line of \unit[0.45]{s}.
Thus, FITC requires per line only \unit[4.7]{\%} of the available
computation time. 

For R2 ($200 \times 200$ pixels), we find the computation time using
FITC to be in total \unit[11.7]{min}.
Following the same argumentation as before for R1, FITC requires only
\unit[6.5]{\%} of the available computation time per line.

%% file: Conclusions.tex
% Paper: SQDM Control with sparse GP
% Section 3: 2DOF Control of SQDM including GPs

\section{Conclusions and Outlook}
\label{sec:conclusions}

For accelerating and improving \SQDM imaging, a two-degree-of-freedom
controller combining Gaussian process regression and extremum seeking
control had been proposed. In this work we have developed this
control paradigm further to make it suitable to be used for arbitrary
surface structures without a priori knowledge. 
To this end, the crucial steps and therewith the main contributions of
this work are, (i) the implementation of an online
hyperparameter optimization and, (ii) the approximation of the Gaussian
process to reduce computation time.

We have shown that with online hyperparameter adaptation it is now possible to
image arbitrary surface structures and found that the fully independent
training conditional (FITC) approximation is the best
working approach for sparsifying the Gaussian process and to reduce the computational
load. The computation time using FITC is sufficiently low for
practical applicability.

Future steps will involve the implementation of the proposed
approach
on the real system and the verification in experiments. Furthermore, future
research will be directed to pixelwise predictions by further lowering
the computation time or to learn the controller directly instead of
separating learning and control as proposed here.